\documentclass[superscriptaddress,twocolumn]{revtex4-1}
\usepackage{graphicx}                   
\usepackage{hyperref}                   
\usepackage{amsmath,amssymb}            
\usepackage{multirow}	
\usepackage{xcolor}	
\usepackage{amsmath,color}

\graphicspath{{s/}}					

\DeclareMathAlphabet{\mathpzc}{OT1}{pzc}{m}{it}

\begin{document}
	
	\author{Azadeh Malekan}
	\affiliation{Department of Physics, University of Tehran, P. O. Box 14395-547, Tehran, Iran}
	\author{Sina Saber}
	\affiliation{Department of Physics, University of Tehran, P. O. Box 14395-547, Tehran, Iran}
	\author{Abbas Ali Saberi}\email{(corresponding author) ab.saberi@ut.ac.ir}
	\affiliation{Department of Physics, University of Tehran, P. O. Box 14395-547, Tehran, Iran}
	\affiliation{Institut f\"ur Theoretische
		Physik, Universit\"at zu K\"oln, Z\"ulpicher Str. 77, 50937 K\"oln,
		Germany}
	
	\title{Exact finite-size scaling for the random-matrix representation of bond percolation on square lattice}

	\begin{abstract}
	We report on the exact treatment of a random-matrix representation of bond percolation model on a square lattice in two dimensions with occupation probability $p$. The percolation problem is mapped onto a random complex matrix composed of two random real-valued matrices of elements $+1$ and $-1$ with probability $p$ and $1-p$, respectively. We find that the onset of percolation transition can be detected by the emergence of power-law divergences due to the coalescence of the first two extreme eigenvalues in the thermodynamic limit. We develop a universal finite-size scaling law that fully characterizes the scaling behavior of the extreme eigenvalue's fluctuation in terms of a set of universal scaling exponents and amplitudes. We make use of the relative entropy as an index of the disparity between two distributions of the first and second-largest extreme eigenvalues, to show that its minimum underlies the scaling framework. Our study may provide an inroad for developing new methods and algorithms with diverse applications in machine learning, complex systems, and statistical physics.  
	\end{abstract}
	\maketitle
	
	\textbf{There exists an exact solution of bond percolation in two dimensions that predicts all critical and divergent properties of the model in the thermodynamic limit i.e., $L\rightarrow\infty$ \cite{sahini1994applications, stauffer2018introduction}.  For more realistic finite systems, however, the standard finite-size scaling hypothesis, formulated for the first time by Fisher \cite{fisher1972scaling, fisher1971theory}, provides a framework to extract various universal features of the system close to the critical occupation threshold $p_c$. When a connectivity percolation is viewed as a random matrix, on the other hand, the model represents a unique correspondence structure with the random matrix theory with novel characteristics \cite{Saber2021Universal}. In the latter case, our present study discovers that the criticality of the system is encoded in the rich scaling properties of the first isolated extreme eigenvalue with Gaussian statistics when approaches the second extreme eigenvalue sitting at the edge of the semicircle law. We present an exact derivation of scaling laws from the coalescence of the two first extremes leading to the formulation of a novel universal finite-size scaling framework which is in some aspects similar to the Family-Vicsek scaling law \cite{vicsek1984dynamic, family1985scaling} emerging in surface-roughness growth of classical systems. We believe that our study opens up promising prospects for the analytical investigation of more complicated interacting systems whose solution has been elusive.  }
	
	Random matrix theory (RMT) has been the subject of various interdisciplinary studies in complex systems and beyond, with a wide range of applications in diverse topics including nuclear and theoretical physics \cite{wigner1951statistical, aaronson2011computational}, number theory\cite{mezzadri2005recent}, statistics\cite{bianchi2011performance}, financial mathematics\cite{pharasi2019complex}, neuroscience\cite{wainrib2013topological}, telecommunications\cite{couillet2011random}, price fluctuations in the stock
	market \cite{plerou1999universal, laloux1999noise}, EEG data of the brain \cite{vseba2003random}, variation of different
	atmospheric parameters \cite{santhanam2001statistics}, and complex networks \cite{jalan2007random}, etc.\\
	Percolation theory, on the other hand, has played an important role in the past four decades to describe and model enormous empirical observations, complex networks, and technological systems experiencing geometrical phase transitions \cite{ saberi2015recent, kalisky2006width, li2021percolation,  fisher1961some,sapoval2004self,vigolo2005experimental,cardy2006power,fortuin1972random,anekal2006dynamic,saberi2010geometrical,knecht2012retention,gruzberg1999exact,endrHodi2014fractality, saberi2020evidence, saberi2013percolation}. 
	
	A common characteristic feature of these two theories is the existence of some specific and popular universality classes that do not depend on fine details of the corresponding models. This indeed provides a framework that accommodates a large number of statistical models and real-world phenomena. In both theories, the statistics of the fluctuations of extremes underlies the universality classes and provides the connection
	with integrable systems. In particular, the statistics of the largest eigenvalue \cite{majumdar2020extreme} and the largest gap in the order parameter \cite{fan2020universal}, determine the system's universality class in RMT and percolation problem, respectively, putting both of them within the extreme-value theory.

	In a recent paper \cite{Saber2021Universal}, the authors have studied a random-matrix realization of a two-dimensional (2D) percolation model. Extensive numerical simulations have unraveled a rich universal framework for the extreme eigenvalues characterized by a set of scaling exponents. The authors have also developed a remarkable finite-size scaling law that governs the universal behavior of the model. Here we aim to provide analytical support for our findings in \cite{Saber2021Universal}, by exact derivation of the scaling exponents and their corresponding amplitudes. A random-matrix realization of a 2D percolation problem with occupation probability $p$ on a square lattice of linear size $L$, was constructed by a $L\times L$ random real matrix $\mathcal{M}'$ whose every element $\mathcal{M}'_{ij}$ is equal to $+1$ with probability $p$ or $-1$ with probability $1-p$. One may notice that the matrix $\mathcal{M}'$ is not equivalent to a 2D bond percolation problem.

	In order to construct a random-matrix representation of a 2D bond percolation model, we define a random complex matrix $\mathcal{M''}=\mathcal{M'}^{h}+i\mathcal{M'}^{v}$, composed of two independent random real matrices $\mathcal{M'}^{h}$ and $\mathcal{M'}^{v}$, as defined above, corresponding to the horizontal and vertical bonds, respectively. $\mathcal{M'}^{h}_{ij}$ and $\mathcal{M'}^{v}_{ij}$ denote as the horizontal and vertical bonds to the right resp. above site ($i, j$). These associate to each open/closed horizontal bond a value $\pm 1$, and to each open/closed vertical bond a complex value $\pm i$. Consequently, every bond-percolation configuration is mapped onto an asymmetric complex matrix $\mathcal{M''}$ with elements $\pm 1\pm i$. We are going to analyze the eigenvalue spectrum of an ensemble of such matrices after symmetrization $\mathcal{M}=(\mathcal{M}''+\mathcal{M}''^\dag)/2$, where $(\cdot)^\dag$ denotes the conjugate transpose of the matrix, to ensure real eigenvalues.
	
Let us first investigate the critical and scaling properties of the extreme eigenvalues in the random-matrix realization of the percolation denoted by the random real matrices \{$\mathcal{M}'$\} with the occupation probability $p$. Our approach can then be straightforwardly extended to the random matrix representation of the 2D bond percolation in terms of the random complex matrices \{$\mathcal{M}''$\}. Our analytical proof indicates that the emergence of the rich finite-size scaling laws when the critical threshold is approached i.e., $p\rightarrow p_c$ \cite{Saber2021Universal}, can be explained based on the emergent deformation that the distribution function of the largest eigenvalue $\lambda_1$ experiences at a certain point ($L^*, p^*$) in the parameter space. This proximity point corresponds to where the distribution of the largest eigenvalue passes from Gaussian ($p>p^*$) to Tracy-Widom ($p<p^*$) distributions. We find that the minimum disparity condition between these two distributions at $p=p^*$, underlies the emergence of the following scaling law: 
\begin{eqnarray}\label{z-exponent}
L^*=A_z|p-p_c|^{-z},
\end{eqnarray}
in which the amplitude $A_z$ and the scaling exponent $z$ seem to be universal. The exact derivation of Equation (\ref{z-exponent}) is one of the main outcomes of the present study.

To sketch the proof for the random real matrices \{$\mathcal{M}'$\}, we use Kullback–Leibler (KL) divergence \cite{kullback1951information,haken2006information}, also known as relative entropy, to quantitatively measure the difference between two probability distributions.
The real symmetric matrix $\mathcal{M}=(\mathcal{M}'+\mathcal{M}'^T)/2$, with $(\cdot)^T$ being the transpose of the matrix, has $\frac{1}{2}L(L+1)$ independent elements $\mathcal{M}_{ij}$ which can take values $-1$, $0$ and $+1$ with probabilities $(1-p)^2$, $2p(1-p)$ and $p^2$, respectively, for $i>j$. The diagonal elements $\mathcal{M}_{ii}$ can be either $-1$ or $+1$ with probabilities $(1-p)$ or $p$, respectively. Every element has the average $\mu=\langle\mathcal{M}_{ij}\rangle=2p-1$ and variance $\sigma^{2}=2 p (1-p)$. The zero mean condition only holds for $p_c=1/2$.	\\
The matrices for all occupation probability $p$ are rescaled by a factor of $1/\sqrt{L}\sigma$ so that the edges of the semicircle distribution function of the bulk eigenvalues would lie in the interval $[-2, 2]$. Since the zero-mean condition holds for $p=1/2$, the distribution of eigenvalues corresponds to the Wigner's semicircle law with edges exactly located at $\pm2$. For $p\ne1/2$, the distribution of eigenvalues  consists of two parts: (i) A semicircle part for the bulk eigenvalues whose edges approach zero as $|p-p_c|\rightarrow 1/2$ with the symmetry $p\leftrightarrow 1-p$ about $p_c$, and (ii) A disjoint isolated distribution of the extremes including the largest (smallest) eigenvalues for $p>1/2$ ($p<1/2$) with the following Gaussian distribution \cite{Saber2021Universal, furedi1981eigenvalues} (without loss of generality, otherwise stated we consider only the case with $p>1/2$): \begin{equation}\label{Gauss}
	\mathcal{P}(\lambda_1)=\dfrac{1}{\sqrt{4\pi\sigma^2}}\exp\left(-\dfrac{\lambda_1-\langle\lambda_{1}\rangle}{4\sigma^2} \right).
\end{equation}
The average largest eigenvalue is given by \cite{ furedi1981eigenvalues}
\begin{equation}
		\begin{aligned} 
		\langle\lambda_{1}\rangle=\mu L+\dfrac{\sigma^{2}}{\mu}, 
		\end{aligned}
		\label{1}
\end{equation} 
which takes the following form for the rescaled matrices by a factor of $1/\sqrt{L}\sigma$ and $p_c=1/2$,
\begin{equation} 
		\begin{aligned}
		\langle\lambda_{1}\rangle=\dfrac{\sqrt{2}L(p-p_c)}{\sqrt{L p (1-p)}}+\dfrac{\sqrt{ L p (1-p)}}{\sqrt{2}L(p-p_c)}.
		\end{aligned}
		\label{2}
\end{equation}

For real symmetric (orthogonal) random matrices, it is known \cite{shcherbina2009edge} that the limiting distribution of the largest eigenvalue sitting at the spectral edge is given by the Gaussian Orthogonal Ensemble (GOE) Tracy-Widom ($\mathcal{TW}_{\beta}$) distribution with Dyson index $\beta=1$ \cite{tracy1994level}. The eigenvalues residing at the semicircle edges are exactly the first extreme eigenvalues $\lambda_1$ for $p=p_c$, while for $p>p_c$ since the first largest eigenvalues get isolated from the bulk within a Gaussian distribution, the second largest eigenvalues $\lambda_2$ are those that reside at the edge with the GOE-$\mathcal{TW}_1$ statistics with the mean $\langle\lambda_2\rangle=2$.

A simple approximation for $\mathcal{TW}_{\beta}$ distribution based on Gamma distribution, has been suggested in \cite{chiani2014distribution} as follows:\begin{equation} 
		\mathcal{TW}_{\beta}(x)\simeq \dfrac{1}{\Gamma(k) \theta^{k}} (x+\alpha)^{k-1}  \exp\left(- \dfrac{x+\alpha}{\theta}\right) , x>-\alpha
		\label{10}\end{equation} 
in which the parameters $k, \theta$, and $\alpha$ are chosen such that to be consistent with the first three moments of $\mathcal{TW}_{\beta}$. For $\mathcal{TW}_{1} $ the parameters are $k=46.4446$, $\theta=0.186054 $ and $\alpha=9.84801$.

Hence, the distribution of the rescaled second largest eigenvalues $\lambda'_2=L^{2/3}(\lambda_{2}-\langle\lambda_{2}\rangle)$ that are residing at the edge of semicircle for $p>p_c$, should be given by
		\begin{equation} 
		\begin{aligned}
	\mathcal{TW}_{1}(\lambda_{2})\simeq \dfrac{L^{2/3}}{\Gamma(k) \theta^{k}}
		(\lambda'_{2}+\alpha)^{k-1}
		\exp\left(- \dfrac{\lambda'_{2}+\alpha}{\theta}\right).
		\end{aligned}
		\label{11}
		\end{equation}

In \cite{Saber2021Universal}, the authors have reported on a transition close to the critical threshold $p_{c} $ when the two Gaussian and GOE distributions merge. In the vicinity of the criticality, the isolated Gaussian part gradually disappears when $p\rightarrow p_c$, and the first largest eigenvalues attach to the edge of  the semicircle. This is where the Gaussian distribution slowly deforms into the GOE-$\mathcal{TW}_1$ distribution. 
To track the difference between these two 
distributions, we use the KL divergence $\mathcal{D}_{\text{KL}}$ to measure the relative entropy from $\mathcal{P}$ to GOE-$\mathcal{TW}_1$,

		\begin{equation} 
		\begin{aligned}
		\mathcal{D}_{\text{KL}}\Big(\mathcal{TW}_1(\lambda) \| \mathcal{P}(\lambda)\Big)=\int_{\lambda_-}^{\lambda_+} d\lambda\mathcal{TW}_1(\lambda) \ln\bigg(\dfrac{\mathcal{TW}_1(\lambda)}{\mathcal{P}(\lambda)}\bigg).
		\end{aligned}
		\label{13}
		\end{equation}
		The upper and lower limits of integration are chosen to include the range of typical fluctuation of $\lambda_2$ about its mean $\langle\lambda_{2}\rangle$, which is of the order of  $\mathcal{O}(L^{-2/3})$, i.e.,
		\begin{equation} 
		\begin{aligned}
		\lambda_\pm=\langle\lambda_{2}\rangle \pm\alpha L^{-\frac{2}{3}}.
		\end{aligned}
		\label{14}
		\end{equation}
The details of integration can be found in the Appendix. After integration, we need to minimize the loss of information in the transition process from the Gaussian to the $\mathcal{TW}_1$, i.e.,
		\begin{equation} 
		\dfrac{d\mathcal{D}_{\text{KL}}}{dp}\bigg|_{L=L^*}=0
		\label{16},
		\end{equation}
		which results in the following condition,
		\begin{equation} 
		\dfrac{L^*(2p-1)}{\sqrt{2 L^* p (1-p)}}+\dfrac{\sqrt{2 L^* p (1-p)}}{L^*(2p-1)}=2,
		\label{17}
		\end{equation}
		with the solution
		\begin{equation} 
		p=\dfrac{1}{2}\pm\dfrac{\sqrt{1+2L^*}}{2(1+2L^*)}.
		\label{18}
		\end{equation}
		By considering $p_c=1/2$, we obtain 
		\begin{equation} 
		p-p_{c}=\pm\dfrac{1}{2\sqrt{1+2L^*}}.
		\label{19}
		\end{equation}
	After taking the limit $ L\rightarrow\infty $, and simple reordering one can find \begin{equation} 
		L^{*}=\frac{1}{8}\vert p-p_{c}\vert^{-2}.
		\label{L-star}
\end{equation}This relation completes our proof for Eq. \ref{z-exponent}, and provides exact predictions for the amplitude $A_z=1/8$ and the scaling exponent $z=2$.

\begin{figure}[t] 
	\centering
	\includegraphics[width=3.4in]{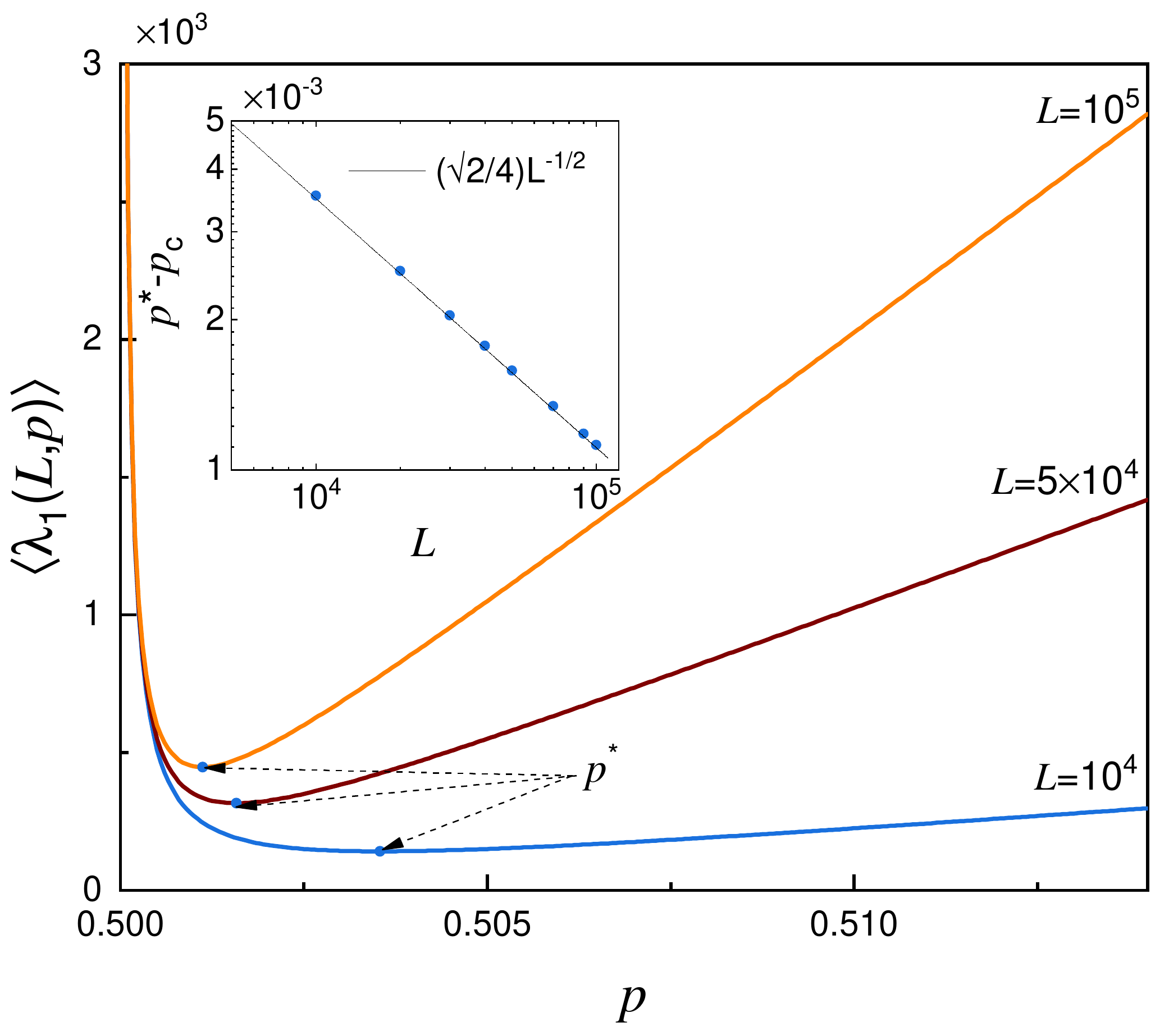}
	\caption{Main: The average largest eigenvalue $\langle\lambda_{1}\rangle$ as function of the occupation probability $p$ very close to $p_c=1/2$ for three different sizes $L=10^4$, $L=5 \times10^4$ and $L=10^5$ (see Eq. (\ref{lambda_mean})). The global minimum $p^*$ of the function (marked by the arrows) approaches to the critical threshold $p^*\rightarrow p_c$ in the infinite size limit $L\rightarrow\infty$. Inset: Power-law decay of $p^*-p_c$ vs $L$ (symbols) which is in perfect agreement with the scaling relation predicted in Eq. (\ref{L-star}) (shown by the solid line) }
	\label{fig1}
\end{figure}

In order to get a clearer understanding of the crossover length scale $L^*$, it is instructive to study the behavior of the average largest eigenvalue as function of $p$ and $L$, i.e., \begin{equation}\label{lambda_mean}
\langle\lambda_{1}(L,p)\rangle=2L(p-p_c)+\dfrac{p(1-p)}{(p-p_c)},\end{equation} written for the original random-matrix realization of percolation problem $\mathcal{M}$ (to be distinguished from Eq. (\ref{2}) which is obtained for a slightly different normalization scheme).
Figure \ref{fig1} shows $\langle\lambda_{1}\rangle$ as a function of $p$ for three different system sizes $L=10^4$, $5\times10^4$ and $10^5$. $\langle\lambda_{1}\rangle$ shows two distinct regimes about the size-dependent global minimum $p_{min}(L)$: for $p>p_{min}$ the first linear term in Eq. (\ref{lambda_mean}) dominates and $\langle\lambda_{1}\rangle$ grows linearly with $p$ with a slope proportional to the system size $L$. For $p<p_{min}$ where $p\rightarrow p_c$, the second term is dominant in Eq. (\ref{lambda_mean}) and $\langle\lambda_{1}\rangle$ diverges to infinity as $p$ approaches to $p_c$. Interestingly, when we demand the following condition to minimize $\langle\lambda_{1}\rangle(p)$,
	\begin{equation}\label{14-}
	\dfrac{d\langle\lambda_{1}\rangle(p)}{dp}\bigg|_{L=L^*}=0;
	\end{equation}
it gives the following relation 
\begin{equation} 
2(L^*-1)=\dfrac{p(1-p)}{(p-p_c)^2}.
\end{equation}
For sufficiently large system sizes $L^*\gg 1$ and  $p_{min}\rightarrow p_c=1/2$ for which $p(1-p)\simeq1/4$, one can recover the same scaling relation as in Eq. (\ref{L-star}) with much lower computational cost (see also the Appendix). This finding indicates that exactly where the average largest eigenvalue is minimized, a minimum disparity is observed between Gaussian and  GOE-$\mathcal{TW}_1$ distributions. Therefore, the two conditions in Eqs. (\ref{16}) and (\ref{14-}) can interchangeably be used to find the scaling relation (\ref{L-star}).

The observations based on numerical simulations in \cite{Saber2021Universal}, have recently revealed that the fluctuation of the average largest eigenvalue about the mean, i.e.,   $\langle\lambda_{1}\rangle_c =\langle\lambda_{1}(L,p)\rangle-2L(p-p_c)$ exhibits a remarkable finite size scaling  \begin{equation}\label{F-V-law}
\langle\lambda_{1}\rangle_c (p, L)=|p-p_c|^{-\alpha} \mathpzc{F}\bigg(\Big(\frac{1}{L}\Big)|p-p_c|^{-z}\bigg),
\end{equation}where the universal function $\mathpzc{F}(x)\rightarrow 1/4$ as $x\rightarrow 0$ (which is now clear from Eq. (\ref{lambda_mean}) with $\alpha=1$ and based on similar arguments above) and $\mathpzc{F}(x)\sim x^{-\alpha/z}$ as $x\rightarrow \infty$, so that $\langle\lambda_{1}\rangle_c$ grows with size like $L^{\alpha/z}$ until it saturates to $(1/4)|p-p_c|^{-\alpha}$ when $L\sim L^*= \frac{1}{8}|p-p_c|^{-z}$. More specifically, 

\begin{equation}\label{F-V-scaling}
\big\langle\lambda_1\big\rangle_c(p, L)\propto
\begin{cases}
|p-p_c|^{-\alpha} & \text{if $L^*\ll L$}\\
(1/L)^{-\beta} & \text{if $L\ll L^*$},\\
\end{cases}       
\end{equation}
where $\alpha$, $\beta$ and $z$ are three positive scaling exponents that characterize the universality class of the largest eigenvalue's fluctuation. However, since both limits in Eq. \ref{F-V-scaling} should meet each other at $L=L^*$, along with Eq. \ref{L-star} imply a scaling relation  $z=\alpha/\beta$ between the exponents.

	We are now in a position to address the universal features of the largest eigenvalues in a random-matrix representation of 2D bond percolation as we have defined earlier.
	To start with, we first build up a random matrix $\mathcal{M''}=\mathcal{M'}^{h}+i\mathcal{M'}^{v}$. The elements of the symmetric matrix   $\mathcal{M}=(\mathcal{M}''+\mathcal{M}''^\dag)/2 $ can take the following possible values with the given occurrence probabilities:
	\begin{equation}\nonumber
		\mathcal{M}_{ij} = \begin{cases}
			+1+i \quad \,\,\,\,\,\,\,\,  \,\,\,\,\,\,\,\,\,\,\, & p^3(1-p) \\
			+1-i \quad \,\,\, \,\,\,\,\,  \,\,\,\,\,\,\,\,\,\, & p^3 (1-p)\\
			-1+i \quad \,\,\, \,\,\,\,\,  \,\,\,\,\,\,\,\,\,\, & p (1-p)^3 \\
			-1-i \quad \,\,\, \,\,\,\,\,  \,\,\,\,\,\,\,\,\,&  p(1-p)^3\\
			+1 \quad \,\,\, \,\,\,\,\,\,\,\,\,\,\,\,\,\,  \,\,\,\,\,\,\,\,\,\,  & p^4+p^2(1-p)^2 \\
			-1 \quad \,\,\, \,\,\,\,\,\,\,\,\,\,\,\,\,\,\,\,\,\,\,\,\,\,\,& (1-p)^4+p^2(1-p)^2\\
			+i \quad \,\,\, \,\,\,\,\,\,\,\,\,\,\,\,\,\,\, \,\,\,\,\,\,\,\,\, & 2 p^2 (1-p)^2 \\
			-i \quad \,\,\, \,\,\,\,\,\,\,\,\,\,\,\,\,\,\,  \,\,\,\,\,\,\,\,\,& 2 p^2 (1-p)^2\\
			0 \quad \,\,\, \,\,\,\,\,\,\,\,\,\,\,\,\,\,\,\,\,  \,\,\,\,\,\,\,\,\,\,& 2 p^3 (1-p)+2 p (1-p)^3.\\
		\end{cases}
		\label{27}
	\end{equation}
Thus, the mean value of the diagonal and off-diagonal elements and their variance are $\mu= 2p-1$ and $\sigma^2= p(1-p)$, respectively. Minimizing  $\mathcal{D}_{\text{KL}}$ according to Eq. (\ref{16}) and repeating the same line of calculations described above, one can find the following scaling relations
	\begin{equation} 
\langle\lambda_{1}\rangle_c=\frac{1}{2}\vert p-p_{c}\vert^{-1}, \hspace{1.0cm} 	L^{*}=\frac{1}{4}\vert p-p_{c}\vert^{-2},	
	\label{28}
\end{equation}
producing the same scaling exponents as for the random-matrix realization of the 2D percolation but with twice the values obtained for the amplitudes.

In conclusion, we present an analytical approach in support of the numerical observations recently reported in \cite{Saber2021Universal} for the random-matrix realization and representation of a 2D bond percolation on square lattice. We show that the scaling relation (\ref{z-exponent}) between the characteristic length scale $L^*$ and the crossover occupation probability $p^*$ emerges at the minimum disparity between the distributions of the disjoint extreme eigenvalues and those residing at the edge of the semicircle law.  We also find that the minimum-disparity condition exactly coincides with the condition where the average extreme eigenvalue of a finite-size system is minimized. This condition provides a much simpler alternative approach for such analytical calculations. 
Based on our scaling arguments, we develop a universal finite-size scaling law (\ref{F-V-law}) which characterizes the whole singular behavior of the model at different regimes on both sides of $L^*$. Usually such finite-size theories are presented in a phenomenological way, however, the scaling law (\ref{F-V-law}) in our study is one of the few relationships of this type that is supported by analytical arguments.

\begin{figure}[t] 
	\centering
	\includegraphics[width=3.4in]{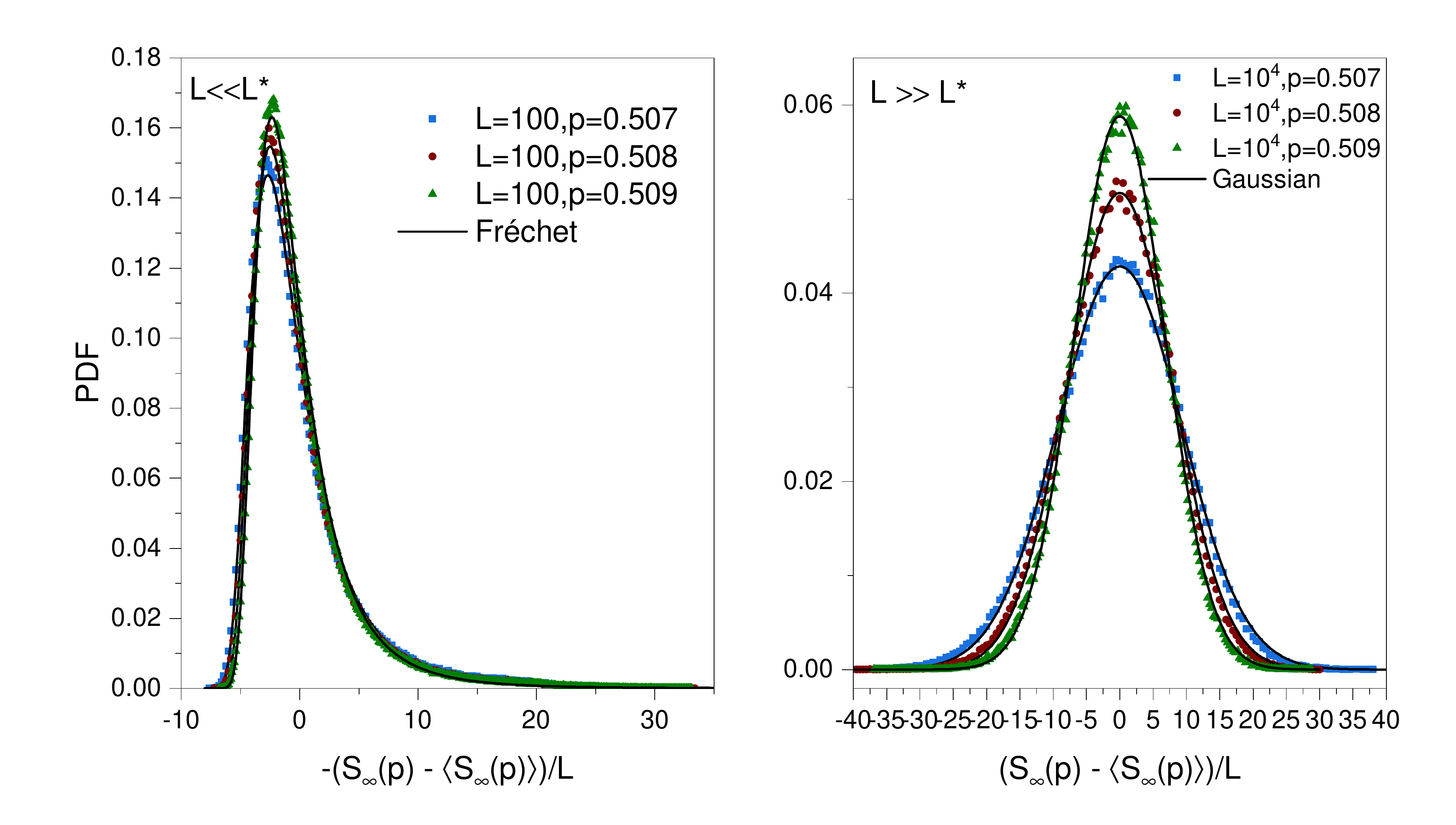}
	\caption{The probability distribution function (PDF) of the scaled fluctuations $(S_\infty(p)-\langle S_\infty(p)\rangle)/L$ of the largest cluster size $S_\infty(p)$ about its average $\langle S_\infty(p)\rangle$ for three different values of the occupation $p=0.507, 0.508$ and $0.509$ close to the critical point $p_c=1/2$ of the bond percolation on a square lattice. Left panel shows the behavior at a small system size $L=10^2\ll L^*$ compatible with the Fr\'echet extreme value distribution (solid lines). The right panel shows the Gaussian statistics of the fluctuations at a large system size $L=10^4$ well above $L^*=A_z|p-p_c|^{-z}$ with $A_z\simeq 4/3$ and $z=43/36$. }
	\label{fig2}
\end{figure}

  An interesting question that may arise is what is the manifestation of the characteristic length scale $L^*$ in the percolation model? As we have seen earlier in this article, the statistical behavior of the largest eigenvalue fluctuations at every occupation probability $p$, changes from a GOE-$\mathcal{TW}_1$ distribution for systems of size $L\ll L^*$ to the Gaussian statistics for $L\gg L^*$. When the critical point is approached $p\rightarrow p_c$, $L^*$ diverges to infinity and thus the GOE-$\mathcal{TW}_1$ statistics governs the criticality of the system. A similar scenario exists for the fluctuations of percolation observables too. As $\lambda_{\text{max}}$ dominates the global behavior of the system in the random matrix representation, the infinite cluster plays the same dominant role in the supercritical percolation regime \cite{stauffer2018introduction}.  This naturally suggests to investigate the scaled fluctuations $(S_\infty(p)-\langle S_\infty(p)\rangle)/L$ of the largest cluster size $S_\infty(p)$ about its average $\langle S_\infty(p)\rangle$ at every given $p>p_c$. We find that there exists a characteristic length scale $L^*\propto |p-p_c|^{-z}$ with $z=\gamma/2=43/36$ about which the distribution of the fluctuations crosses over from a Fr\'echet extreme value distribution for systems of size $L\ll L^*$ to the Gaussian statistics for $L\gg L^*$. The results are shown in Fig. \ref{fig2}. Similar to the thermal phase transitions where the fluctuations of the order parameter are related to the susceptibility $\chi$, it can be shown  \cite{coniglio1980fluctuations} that an analogous relation holds for the percolation model. With this analogy, it is straightforward to check that $\langle|S_\infty(p)-\langle S_\infty(p)\rangle|\rangle/L\propto \sqrt{\chi}\sim (p-p_c)^{-\gamma/2}$. However, the exact proof of our above numerical observation for the scaling exponent $z=\gamma/2=43/36$ calls for further future research. The mathematical tools already developed in the random matrix theory enabled us to sketch the proof for the largest eigenvalue fluctuations, and we hope our work motivates future studies for the analytical investigation of the emergence of extreme value statistics in general percolation models.

\section*{Acknowledgments}
We would like to thank S.N. Majumdar, H. Spohn, and especially P.L. Ferrari for useful discussions related to this work. We also thank the High-Performance Computing (HPC) center at the University of Cologne, Germany,
where a part of related computations has been carried out.

\section*{AUTHOR DECLARATIONS}
\subsection*{Conflict of Interest}
The authors have no conflicts to disclose.

\section*{DATA AVAILABILITY}
The data that support the findings of this study are available
from the corresponding author upon request.


\section*{Appendix} \label{six}	
\appendix
\setcounter{equation}{0}
 \renewcommand{\theequation}{A-\arabic{equation}}

This Appendix presents the details of integration of $ \mathcal{D}_{\text{KL}} $ given in Eq. (\ref{13}) which resulted in Eq. (\ref{17}). \\
In the infinite size limit $L\rightarrow\infty$ the average second largest eigenvalue  $\langle\lambda_{2}\rangle$
for $p>1/2$ resides exactly at the edge of the semicircle law. However, for a system of finite-size $L$, it is located at an average distance of order $(k\theta -\alpha)/L^{2/3}$ from the edge of the semicircle law \cite{chiani2014distribution}. Considering this shift, Equation (\ref{11}) can be rewritten as follows:
		\begin{equation} 
		\begin{aligned}
		\mathcal{TW}_{1}(\lambda)\simeq\frac{L^{2/3}}{\Gamma(k) \theta^{k}}&\left( L^{2/3}(\lambda - \langle\lambda_{2}\rangle+\dfrac{k \theta -\alpha}{L^{2/3}})+\alpha\right) ^{k-1}\\
		\times\exp&\left(- \frac{L^{2/3}(\lambda-\langle\lambda_{2}\rangle+\frac{k \theta -\alpha}{L^{2/3}})+\alpha}{\theta}\right).
		\end{aligned}
		\label{A0}
		\end{equation}
Considering this function along with $\mathcal{P}(\lambda)$ given in Eq. (\ref{Gauss}), one can write down the logarithm in the integrand of Eq. (\ref{13}) as follows,
\begin{equation} 
		\begin{aligned}
	\ln\bigg(\dfrac{\mathcal{TW}_1(\lambda)}{\mathcal{P}(\lambda)}\bigg)&=\ln\dfrac{\sqrt{4\pi\sigma^{2}} L^{\frac{2}{3}}}{\Gamma(k)\theta^{k}}\\
		&+(k-1) \ln\left(L^{2/3}(\lambda-\langle \lambda_{2}\rangle+\dfrac{k \theta -\alpha}{L^{2/3}})+\alpha\right)\\ 
		&-\dfrac{L^{2/3}(\lambda-\langle\lambda_{2}\rangle+\dfrac{k\theta -\alpha}{L^{2/3}})+\alpha}{\theta}\\
		&+\dfrac{\sigma^2(\lambda-\langle\lambda_{1}\rangle)^2}{4}
		\end{aligned}
		\label{A2}
		\end{equation}
	with $ k=46.4446$, $\theta=0.186054 $ and $ \alpha=9.84801 $. 
Thus the integral $\mathcal{D}_{KL}$ in Eq. (\ref{13}) can be decomposed into four sub-integrals,
\begin{equation} 
		\mathcal{D}_{KL}=I_{1}+I_{2}+I_{3}+I_{4}, 
		\label{A3}\end{equation}
with the following individual forms
	\begin{equation} 
		\begin{aligned}
	I_{1}=C_1\int_{\lambda_-}^{\lambda_+}\left(L^{2/3}(\lambda-\langle\lambda_{2}\rangle +\frac{k\theta -\alpha}{L^{2/3}})+\alpha\right)^{k-1}\\
		\times\exp\left(-\frac{L^{2/3}(\lambda-\langle\lambda_{2}\rangle +\frac{k\theta-\alpha}{L^{2/3}})+\alpha}{\theta}\right)  d\lambda\\
		I_{2}=C_2\int_{\lambda_-}^{\lambda_+} \ln\left(L^{2/3}(\lambda-\langle\lambda_{2}\rangle+\frac{k \theta -\alpha}{L^{2/3}})+\alpha \right)\\
		\times\left( L^{2/3}(\lambda-\langle\lambda_{2}\rangle +\frac{k \theta -\alpha}{L^{2/3}})+\alpha\right)^{k-1} \\
		\times\exp\left( -\frac{L^{2/3}(\lambda-\langle\lambda_{2}\rangle+\frac{k \theta -\alpha}{L^{2/3}})+\alpha}{\theta}\right) d\lambda\\
		I_{3}=-C_3\int_{\lambda_-}^{\lambda_+}\left(\dfrac{L^{2/3}(\lambda-\langle\lambda_{2}\rangle+\frac{k\theta-\alpha}{L^{2/3}})+\alpha}{\theta}\right)\\
		\times\left( L^{2/3}(\lambda-\langle \lambda_{2}\rangle+\dfrac{k \theta -\alpha}{L^{2/3}})+\alpha\right) ^{k-1}\\
		\times\exp\left(-\frac{L^{2/3}(\lambda-\langle\lambda_{2}\rangle+\frac{k \theta -\alpha}{L^{2/3}})+\alpha}{\theta}\right) d\lambda\\
		I_{4}=\dfrac{C_3}{4}\int_{\lambda_-}^{\lambda_+}
		\sigma^2 (\lambda-\langle\lambda_{1}\rangle) ^{2} \left( L^{2/3}(\lambda-\langle\lambda_{2}\rangle+\frac{k\theta -\alpha}{L^{2/3}})+\alpha\right) ^{k-1}\\ 
		\times\exp\left(- \frac{L^{2/3}(\lambda-\langle\lambda_{2}\rangle +\frac{k \theta -\alpha}{L^{2/3}})+\alpha}{\theta} \right) d\lambda,
		\end{aligned}
		\label{A7}
		\end{equation}
where $C_1$, $C_2$, and $C_3$ are defined as follows\begin{equation} 
		\begin{aligned}
		&C_1=\dfrac{L^{2/3}}{\Gamma(k) \theta^{k}}\,\ln\dfrac{\sqrt{4 \pi \sigma^{2}} L^{\frac{2}{3}}}{\Gamma(k) \theta^{k}}, \\
		&C_2=(k-1)\dfrac{L^{2/3}}{\Gamma(k) \theta^{k}},\\
		&C_3=\dfrac{L^{2/3}}{\Gamma(k) \theta^{k}}.
		\label{A10}
		\end{aligned}
		\end{equation}
The lower and upper limits of integration in (\ref{13})  can be given as follows by considering the finite size effects, \begin{equation}
		\begin{aligned} 
	    \lambda_-=& \langle\lambda_{2}\rangle-(k \theta -\alpha)L^{-\frac{2}{3}}-(\alpha-\varepsilon) L^{-\frac{2}{3}},\\
		\lambda_+=& \langle\lambda_{2}\rangle-(k \theta -\alpha)L^{-\frac{2}{3}}+(\alpha-\varepsilon) L^{-\frac{2}{3}},
		\end{aligned}
		\label{A12}
		\end{equation}
in which the parameter $ \varepsilon $ is defined to avoid the singularity in the lower limit of the integration. After integration, one can find	\begin{equation} 
		I_{1}=C_1 \theta^{-k} L^{-\frac{2}{3}} \left( \Gamma(k, \frac{2\alpha-\varepsilon}{\theta})-\Gamma(k,\varepsilon)\right), 
		\label{A13}
		\end{equation}
		\begin{equation} 
		\begin{aligned}
		I_{2}=& -\dfrac{C_2 (2\alpha-\varepsilon)^k}{ L^{\frac{2}{3}}k^{2}}\,_{2}F_{1}(k,k;1+k,1+k;\frac{-2\alpha+\varepsilon}{\theta})\\
		&+\dfrac{C_2 \varepsilon^k}{L^{\frac{2}{3}}k^{2}}\,\,_{2}F_{1}(k,k;1+k,1+k;\frac{-\varepsilon}{\theta})\\
		&+\dfrac{C_2 (2\alpha-\varepsilon)^k \theta^{k}ln\frac{2\alpha-\varepsilon}{\theta}}{ L^{\frac{2}{3}} k}\left( \Gamma(1+k)-k\Gamma(k,\frac{2\alpha-\varepsilon}{\theta})\right)\\
		&- \dfrac{C_2 \varepsilon^k \theta^{k} ln\frac{\varepsilon}{\theta}}{ L^{\frac{2}{3}} k} \left(  \Gamma(1+k)-k\Gamma(k,\frac{\varepsilon}{\theta})\right),
		\end{aligned}
		\label{A14}
		\end{equation}
	with $_{2}F_{1}$ being the hypergeometric function. The integration of the third term gives
\begin{equation} 
		I_{3}=C_3 L^{\frac{-2}{3}} \theta^{-k} \left( \Gamma(1+k,\frac{2\alpha-\varepsilon}{\theta})-\Gamma(1+k,\varepsilon)\right). 		\label{A15}
		\end{equation}
The dependence on $p$ only appears in $I_4$ through $\langle\lambda_{1}\rangle$, 
\begin{equation} 
		\begin{aligned}
		I_{4}=&\dfrac{-C_3}{ L^{2} \theta^{k}} \left(L^{\frac{2}{3}}(\langle\lambda_1\rangle-\langle\lambda_2\rangle+\frac{k \theta -\alpha}{L^{2/3}})+\alpha\right)^2\\
		&\times\left(\Gamma(k,\frac{2\alpha-\varepsilon}{\theta})-\Gamma(k,\varepsilon)\right)\\
		&+\dfrac{2\theta C_3}{ L^{2} \theta^{k}}\left( L^{\frac{2}{3}}(\langle\lambda_1\rangle-\langle\lambda_2\rangle+\frac{k \theta-\alpha}{L^{2/3}})+\alpha\right)
		\\
		&\times\left(\Gamma(1+k,\frac{2\alpha-\varepsilon}{\theta}) -\Gamma(1+k,\varepsilon)\right)\\
		&-\dfrac{ \theta^{2} C_3}{ L^{2} \theta^{k}} \left( \Gamma(2+k,\frac{2\alpha-\varepsilon}{\theta})-\Gamma(2+k,\varepsilon)\right).
		\end{aligned}
		\label{A16}
		\end{equation}
The minimum-disparity condition then requires,
\begin{equation} \nonumber
\dfrac{d\mathcal{D}_{\text{KL}}}{dp}\bigg|_{L=L^*}=0,
\end{equation}
which receives contributions only from $I_4$,
\begin{equation} 
		\dfrac{d\mathcal{D}_{KL}}{d\langle\lambda_1\rangle}\dfrac{d\langle\lambda_1\rangle}{dp}\bigg|_{L=L^*}=\dfrac{d I_{4}}{d\langle\lambda_1\rangle} \dfrac{d\langle\lambda_1\rangle}{dp}\bigg|_{L=L^*}=0.
		\label{A17}
		\end{equation}
For the above relation to hold, either $dI_4/d\langle\lambda_1\rangle$	or $d\langle\lambda_1\rangle/dp$ should vanish. 
Therefore, the first condition can be read off as \begin{equation} 
		\begin{aligned}
		\dfrac{d I_{4}}{d\langle\lambda_1\rangle}=& 2L^{2/3} \left( L^{2/3}( \langle\lambda_1\rangle-\langle\lambda_2\rangle+\frac{k \theta -\alpha}{L^{2/3}})+\alpha\right)\\
		&\times\left( \Gamma(k,\frac{2\alpha-\varepsilon}{\theta})-\Gamma(k,\varepsilon)\right) \\
		&-2\theta L^{2/3} \left( \Gamma(1+k,\frac{2\alpha-\varepsilon}{\theta})-\Gamma(1+k,\varepsilon)\right)\bigg|_{L=L^*}=0
		\end{aligned}
		\label{A19}
		\end{equation}
For small $\varepsilon$ one can make use of the following approximations for the Gamma functions,
		\begin{equation} 
		\begin{aligned}
		&\Gamma(k,\frac{2\alpha-\varepsilon}{\theta})-\Gamma(k,\varepsilon)\simeq -\Gamma(k,\varepsilon)\\
		&\Gamma(1+k,\frac{2\alpha-\varepsilon}{\theta})-\Gamma(1+k,\varepsilon)\simeq -k \Gamma(k,\varepsilon)\\
		\end{aligned}
		\label{A20}
		\end{equation}
Substituting (\ref{A20}) into (\ref{A19}), one can find that 
\begin{equation}	\langle\lambda_1\rangle=\langle \lambda_2\rangle.\end{equation}
Using Eq. (\ref{2}) one can find \begin{equation} 
	\dfrac{L^*(2p-1)}{\sqrt{2 L^* p (1-p)}}+\dfrac{\sqrt{2 L^* p (1-p)}}{L^*(2p-1)}=2,
	\label{A23}
	\end{equation}	which is our Eq. (\ref{17}) in the main text.
	The second condition \begin{equation}\nonumber
	\dfrac{d\langle\lambda_{1}\rangle(p)}{dp}\bigg|_{L=L^*}=0,
	\end{equation}
	is thoroughly discussed in the text.
		
		
\bigskip		
		\bibliography{refs} 

\end{document}